\documentclass[twocolumn,12pt,cites,graphicx]{article}
\usepackage{epsfig}

\title{Orientationally ordered aggregates of stiff polyelectrolytes in the presence of multivalent salt}
\author{Sarah Mohammadinejad\\
Institute for Advanced Studies in Basic Sciences (IASBS), P. O. Box
45195-1159,\\ Zanjan 45195, Iran. E-mail: sarah@iasbs.ac.ir
\and Hossein Fazli \\
Institute for Advanced Studies in Basic Sciences (IASBS), P. O. Box
45195-1159,\\ Zanjan 45195, Iran. E-mail: fazli@iasbs.ac.ir
\and Ramin Golestanian \\
Department of Physics and Astronomy, University of Sheffield,
\\Sheffield S3 7RH, UK. E-mail: r.golestanian@sheffield.ac.uk}
\date{Received XXXXth Month, 200X\\Accepted XXXXth Month, 200X\\DOI: 10.1039/}
\begin{document}

\maketitle
\renewcommand{\thefootnote}{\fnsymbol{footnote}}

\noindent Aggregation of stiff polyelectrolytes in solution and
angle- and distance-dependent potential of mean force between two
like-charged rods are studied in the presence of 3-valent salt using
molecular dynamics simulations. In the bulk solution, formation of
long-lived metastable structures with similarities to the raft-like
structures of actin filaments is observed within a range of salt
concentration. The system finally goes to a state with lower free
energy in which finite-sized bundles of parallel polyelectrolytes
form. Preferred angle and interaction type between two like-charged
rods at different separations and salt concentrations are also
studied, which shed some light on the formation of orientationally
ordered structures.

\section{Introduction}
\label{intro}It is well known that highly charged anionic biological
polyelectrolytes such as DNA and filamentous actin (F-actin) could
attract each other in the presence of multivalent cations
\cite{electrostatics}. Attraction between similarly charged
polyelectrolytes, which is due to electrostatic correlations and can
not be described in the framework of mean field theories such as
Poisson-Boltzmann formalism, has been the subject of intense ongoing
investigations in the past few decades
\cite{electrostatics,oosawa,guldbrand,israelachvili,ha,gronbech,stevens,rg,shklovskii,bloomfield,tang,podgornik,naji}.
Aggregation of stiff polyelectrolytes in solution in the presence of
multivalent salt arises from such like-charge attractions.

Depending on their different roles in the cell function, actin
filaments assemble into networks (large crossing angle) or bundles
(nearly parallel), respectively, helped by bundling or cross-linking
proteins \cite{lodish}. A similar pattern---namely, formation of
both networks and dense bundles---has also been observed
experimentally in solutions of F-actin rods with multivalent salt,
depending on the salt concentration \cite{G.Wong}. In a particular
range of salt concentrations, a multi-axial liquid crystalline phase
of actin rods has been observed in which aggregates resembling
stacks of raft-like structures form. These structures, which consist
of a combination of mutually parallel and perpendicular F-actin
rods, change to bundles of parallel rods with increasing multivalent
salt concentration \cite{G.Wong}. This is a manifestation of the
rich complexity of high density aggregates of linear
polyelectrolytes due to {\em frustration} in accommodating strong
electrostatic interactions and geometrical constraints
simultaneously. Similar effects arising from electrostatic
frustration have been studied theoretically in the context of
rodlike polyelectrolyte aggregates \cite{HF1}, rodlike
polyelectrolyte brushes \cite{HF2}, and star polyelectrolytes
\cite{likos}.

Formation of novel raft-like aggregates of actin rods in the
presence of multivalent salt has been studied both theoretically and
using computer simulation methods. As a model system, nematic
mixture of rigid rods and strong $\frac{\pi}{2}$ cross-linkers has
been studied theoretically using Onsager excluded volume theory, and
the possibility of formation of stacks of raft-like structures has
been shown \cite{borukhov1}. Distance-dependent interaction between
two parallel similarly charged rigid rods in the presence of
explicit mono- and multivalent salt ions and angle-dependent
interaction between them at close center-to-center separation have
been studied by Lee {\em et al.} using molecular dynamics (MD)
simulations \cite{Lee}. In their study, Lee {\em et al.} have shown
that just above the threshold value of multivalent salt
concentration needed for attraction between parallel rods, the
preferred angle at close separations is $90^\circ$, and they have
concluded that this is in agreement with formation of raft-like
structures in multivalent salt solution of F-actin rods.

The phase behavior of a model system composed of similarly charged
rigid rods and inter-rod linkers has also been studied using
Debye-H\"{u}ckel theory of electrostatic interaction between the
rods and considering the role of the linkers in the effective
inter-rod interaction potential \cite{borukhov2}. In this study, it
has been shown that aggregates with different orientational
orderings of the rods can be formed at different values of rod and
linker concentrations. Recently, aggregation of stiff
polyelectrolytes has been studied using MD simulations in the
presence of multivalent counterions \cite{Sayar} and in the
multivalent salt solution \cite{Fazli}. In Ref. \cite{Fazli} it is
shown that the dominant kinetic mode in aggregation process is the
case of one end of one rod meeting others at right angle with no
appreciable energy barrier and the polyelectrolytes have a tendency
to form finite-sized bundles. The kinetics of aggregation and bundle
formation of actin has also been recently probed experimentally
\cite{gerard-ramin}. Using two different fluorescently labeled
populations of F-actin, it was shown that the growth mode of actin
bundles in the late stages is predominantly longitudinal, and that
the bundles can freely exchange filaments with the solution and
therefore are not in a frozen state \cite{gerard-ramin}. It was also
found that the energy barrier for the aggregation process is
negligibly small (of the order of $k_{\rm B} T$) \cite{gerard-ramin}
in agreement with the theoretical result of Ref. \cite{Fazli}.

Considering the complexity of the problem of highly charged rod-like
polyelectrolytes at high densities, it seems that more studies are
needed to fully unravel the dynamic behavior of the system. Here we
use a combination of simulation of the bulk solution of stiff
polyelectrolytes and calculation of the effective interaction
potential between a pair of polyelectrolytes in the presence of
multivalent salt to better understand the orientational ordering of
the polyelectrolytes in the aggregation process. We use MD
simulations to study stiff polyelectrolytes in the bulk solution in
the presence of explicit counter-ions and ions of multivalent salt.
We also examine the angle- and distance-dependent potential of mean
force between two similarly charged rigid rods in the presence of
multivalent salt to help understand the observations of the bulk
solution simulations. We find that there could be {\em three}
different regimes in the aggregation behavior of stiff
polyelectrolytes in a multivalent salt solution, depending the salt
concentration. When the salt concentration is low, the system does
not undergo aggregation because the attraction between
polyelectrolytes is not strong enough to trigger it. In an
intermediate range of salt concentrations, relatively long-lived
structures with special orientational ordering of polyelectrolytes
form. In these scaffold-like structures, polyelectrolytes are
predominantly joined either parallel or perpendicular to each other
(see Fig. \ref{fig:snapshot1}), which is presumably due to the
dominant kinetic mode of aggregation \cite{Fazli}. After
sufficiently long time, however, mutually perpendicular rods in each
structure slide against each other and join up in parallel, such
that eventually bundles of parallel polyelectrolytes form. These
meta-stable structures have similarities to the raft-like structures
of actin filaments observed experimentally \cite{G.Wong}. In the
high salt concentration regime, the polyelectrolytes directly
aggregate into bundles of parallel rods and do not exhibit
right-angle configurations in their kinetic pathway of aggregation.

We also calculate the angle- and distance-dependent potential of
mean force between two similarly charged rigid rods, and find that
in the low salt concentration regime there is no attraction between
rods. In the regime with intermediate values of salt concentration,
the interaction between parallel rods is attractive but in a range
of center-to-center separation where the preferred angle between the
rods is $90^\circ$. In this regime, the parallel configuration is
preferred only at very close separations. In the high salt
concentration regime, the parallel configuration is preferred at all
separations and the interaction between parallel rods is attractive.

We also check the stability (life-time) of raft-like structures by
constructing such structures and monitoring their subsequent
evolution as shown in Fig. \ref{fig:snapshot2}, at different values
of salt concentration. We find that these structures are not stable
in the low and high salt concentration regimes: while at low salt
concentrations mutual repulsion between rods destroys such
structures, at high salt concentrations mutually perpendicular rods
slide against each other in a very short time and the raft-like
structures change to bundles of parallel rods. In the intermediate
salt concentration regime, raft-like structures are relatively
long-lived. We have observed that the life time of such structures
rapidly increases with the number of rods.

The rest of the paper is organized as follows. In Sec.
\ref{sec:Agg}, the simulation method and the results of our studies
on the aggregation of polyelectrolytes are explained. Section
\ref{sec:Pot} is devoted to the studies on the potential of mean
force between rods at various separations and angles, while Sec.
\ref{sec:Stb} describes the studies on the stability of raft-like
structures. Finally, Sec. \ref{sec:Con} concludes the paper.

\section{Aggregation of stiff polyelectrolytes in the bulk solution}
\label{sec:Agg}

\subsection{The model and the simulation method}
\label{subsec:model_method2}

In our simulations, which are performed with MD simulation package
ESPResSo v.1.8 \cite{Espresso}, each polyelectrolyte rod is composed
of $N_m=21$ spherical charged monomers of charge $-e$ ($e$ is
electronic charge) and diameter $\sigma$. We have considered $N_p$
polyelectrolytes with their monomers being bonded to each other via
a FENE \cite{Grest} (finite extensible non linear elastic)
potential, with a the separation between them being fixed at
$a=1.1\sigma$. The bending rigidity of polyelectrolyte chains is
modeled with bond angle potential $U_{\phi}=k_{\phi}(1-\cos{\phi})$
with $k_{\phi}=400k_{\rm B}T$ in which $\phi$ is the angle between
two successive bond vectors along the polyelectrolyte chain. We use
$N_c=N_p\times N_m$ monovalent counter-ions with charge $+e$ to
neutralize the polyelectrolyte charges. We also model 3-valent salt
as $N_{s+}$ positive ions with charge $+3e$ and $N_{s-}=3 N_{s+}$
negative ions with charge $-e$. We follow Ref. \cite{Lee} for the
definition of salt concentration: $c_{3:1} \equiv N_{s-}/N_c$ (for
example $c_{3:1}=0.5$ means that $N_{s-}=0.5\times N_c$). The
monomer number density in our simulations is $0.04\;\sigma^{-3}$,
which is about two times the concentration we have used in Ref.
\cite{Fazli}. In addition to long-range Coulomb interaction, we
include short-range Lennard-Jones repulsion between particles, which
adds an energy scale $\epsilon$ and a length scale $\sigma$ to the
system. MD time step in our simulations is $\tau=0.01\;\tau_0$ in
which $\tau_0=\sqrt{\frac{m\sigma^2}{\epsilon}}$ is the MD time
scale and $m$ is the mass of the particles. The temperature is fixed
at $k_{\rm B}T=1.2\epsilon$ using a Langevin thermostat. We use the
P3M method to apply periodic boundary conditions for long-range
Coulomb interaction in the system. The strength of the electrostatic
interaction energy relative to the thermal energy can be quantified
by definition of the Bjerrum length $l_{\rm
B}=\frac{e^2}{\varepsilon k_{\rm B}T}$, where $\varepsilon$ is the
dielectric constant of the solvent and in our simulations $l_{\rm
B}=3.2\sigma$. In water at room temperature $l_{\rm B}=7\AA$ and the
value of $l_{\rm B}=3.2\sigma$ in our simulations means that
$\sigma=2.2 \AA$ and the separation between charged monomers is
$a=2.5\AA$. By using the relation for the effective viscosity of the
solvent $\eta\simeq 2.4 \sqrt{m\epsilon}/\sigma^2$ (from Ref.
\cite{Dunweg}), we can deduce an estimate for the microscopic time
$\tau_0 \simeq \eta \sigma^3/(2 k_{\rm B} T)$. Using the viscosity
of water and room temperature, we find $\tau_0=1.3$ ps.

In the beginning of bulk solution simulations, we fix the
polyelectrolytes in space and leave all the ions to fluctuate around
them for 100000 MD time steps, which is enough to equilibrate the
ions for every given configuration of the polyelectrolytes. After
equilibration, we remove the constraint on the polyelectrolytes and
study the aggregation process in the system.

\subsection{Results}
\label{subsec:Results2}

\begin{figure}[ht]
\begin{center}
\resizebox{1.99\columnwidth}{!}{
\includegraphics{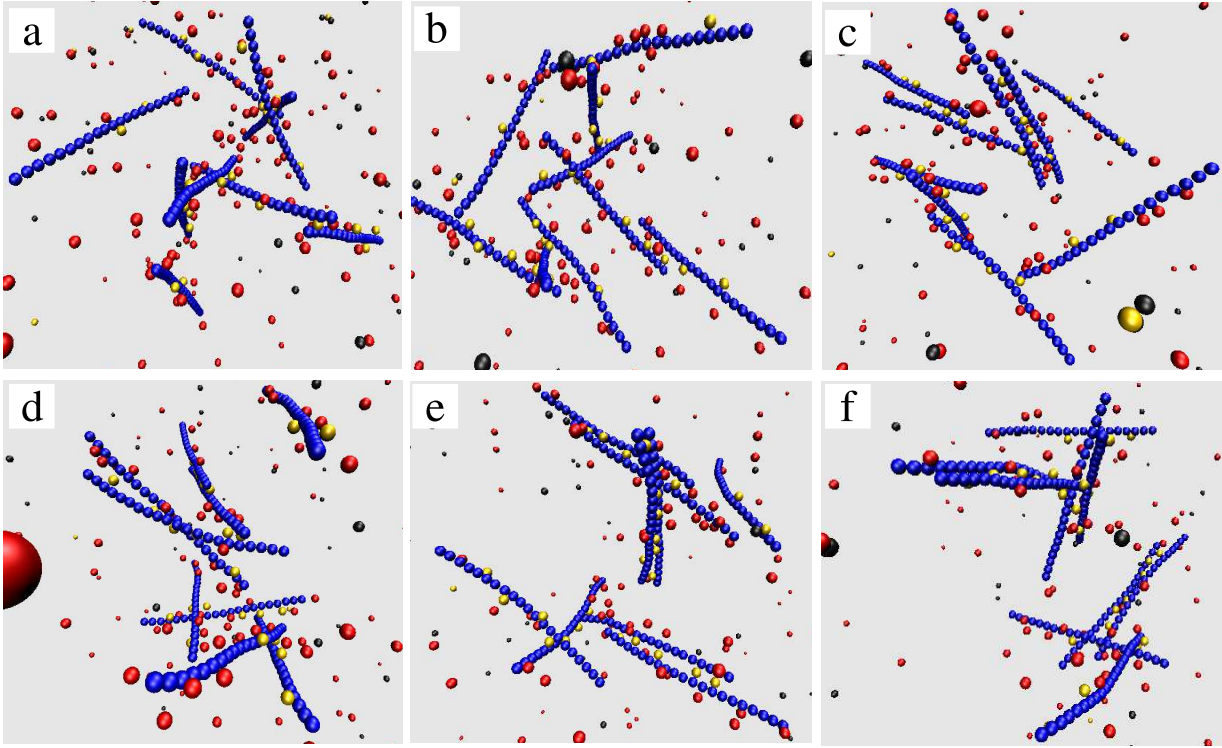}}
\caption{Snapshots of the system containing $N_p=9$ polyelectrolytes
in solution with 3-valent salt concentration $c_{3:1}=0.5$ as a
function of time. +3 salt ions are shown by golden (light) spheres,
+1 counter-ions by red (dark), and -1 salt ions by gray spheres.
Corresponding times of snapshots in units of $\tau_0$: (a) 10, (b)
25, (c) 40, (d) 55, (e) 70, (f) 85. Aggregates of polyelectrolytes
with scaffold-like structures can be seen in the snapshots e and f.
Finally, all of the polyelectrolytes form a single bundle (last
configuration is not shown here).} \label{fig:snapshot1}
\end{center}
\end{figure}

Our bulk solution simulations show that with changing multivalent
salt concentration, three different regimes can exist. At salt
concentrations less than a lower value $c_{3:1}\simeq 0.2$ no
aggregation is observed in the system. In this regime, the salt
concentration is lower than its minimum value needed for like-charge
attraction between polyelectrolytes. When a 3-valent salt ion links
two monomers from two different polyelectrolytes to each other,
thermal fluctuations separate them again and the system does not
undergo an aggregation process. At intermediate values of the salt
concentration $0.2\leq c_{3:1}\leq 1.2$, aggregation of
polyelectrolytes can be observed and the dominant kinetic mode of
aggregation is that of one end of one polyelectrolyte meeting others
at a right angle, as previously reported \cite{Fazli}. During the
aggregation process in this regime, long-lived structures of
polyelectrolytes form, and in each structure polyelectrolytes are
dominantly joined either parallel or perpendicular to each other
(see Fig. \ref{fig:snapshot1} f).

In this regime, after sufficiently long time mutually perpendicular
polyelectrolytes in each scaffold-like structure slide against each
other and finally form bundles of parallel polyelectrolytes. The
kinetic mode of evolution from scaffold-like structures to bundles
is considerably slower than the typical kinetic modes of the system.
For example, in a system containing nine polyelectrolytes, the time
that takes for the scaffold-like structure (Fig. \ref{fig:snapshot1}
f) to evolve into the final configuration of a collapsed parallel
bundle is greater than the time interval between snapshots $a$ and
$e$ of Fig. \ref{fig:snapshot1}, by a factor of at least six. This
suggests that the formation of long-lived scaffold-like structures
corresponds to meta-stable states of the system and the states with
bundles of parallel polyelectrolytes have lower free energies. We
have checked that the formation of the meta-stable structures is
independent of the random number sequence used in the simulation, as
well as the initial configuration of the polyelectrolytes. In other
words, in the intermediate salt concentration range the relaxation
pathway to equilibrium appears to always involve intermediate
meta-stable structures. These structures have similarities to the
raft-like structures of actin filaments observed experimentally
\cite{G.Wong} (see below for a discussion).

In the regime with high values of salt concentration ($c_{3:1}\geq
1.2$), we observe a different kinetic mode of aggregation in the
system. In this regime, the polyelectrolytes directly aggregate into
bundles of parallel polyelectrolytes and the salt concentration
washes out the right angle configuration from the kinetic pathway of
aggregation. The aggregation process in this regime is considerably
faster than that of the regime with intermediate salt concentration.

To shed some light on the behavior of the system and better
understand these results, we study the potential of mean force
between two polyelectrolyte rods.

\section{Potential of mean force between charged rods}
\label{sec:Pot}
\subsection{The model and the simulation method}
\label{subsecsec:model_method3}

In our simulations of two similarly charged rigid rods we consider
two rods, each composed of $N_m=21$ similarly charged spherical
monomers of diameter $\sigma$ and charge $-e$, $N_c=2N_m$
counterions of charge $+e$, and 3-valent salt of concentration
$c_{3:1}$. The monomer number density in these simulations is
$10^{-4}\;\sigma^{-3}$, which corresponds to a simulation box of
length $L \approx 75 \sigma$ that is large enough to avoid finite
size effects. The rods are fixed on two parallel plates of
separation $R$ and the angle between the rods is $\theta$ (see the
schematic configuration in the inset of the middle part of Fig.
\ref{fig:deltau_r_theta}). For each set of values of $R$, $\theta$,
and $c_{3:1}$, the two rods are fixed during the simulation and only
the counterions and the salt ions are free to move. After
equilibration of the system we calculate the average force on each
monomer and obtain the total force and the total torque around the
center-to-center line ($y$-axis) for one of the rods. We integrate
the force with respect to the center-to-center distance, $R$, to
obtain the $R$-dependence of the potential of mean force, and
similarly integrate the torque with respect to the angle $\theta$ to
obtain its $\theta$-dependence (corresponding to rotation around
$y$-axis). We set the zero of the potential of mean force at $R=7.5
\sigma$ for obtaining its $R$-dependence, and at $\theta=90^\circ$
in the calculation of its $\theta$-dependence.

\subsection{Results}
\label{subsec:Results3}

We first repeat the calculation of the potential of mean force
between two parallel rods at different values of salt concentration
and center-to-center separation as in Ref. \cite{Lee}. Our results
show that for 3-valent salt concentrations higher than
$c_{3:1}=0.2$, there is attraction between parallel rods (see Fig.
\ref{fig:deltau_r_parallel}), in agreement with the results of Ref.
\cite{Lee}. We now consider the following question: if one allows
the angle between the rods to vary for other values of $R$, will the
rods prefer to be parallel? To answer this question, we calculate
the angle-dependent potential of mean force at different values of
$R$. We find that at the lowest value of $R$ in our simulations,
$R=2.5\sigma$, the preferred angle at salt concentrations less than
a threshold value $c_{3:1}\simeq0.2$ is $\theta=90^\circ$, and at
higher salt concentrations, the parallel configuration is preferred
(see $R=2.5\sigma$ part of Fig. \ref{fig:deltau_r_theta}). While
this is the same result as reported in Ref. \cite{Lee}, we find that
upon increasing $R$ (from $R=2.5\sigma$), the threshold value of
salt concentration below which $\theta=90^\circ$ is preferred
increases (see the parts of Fig. \ref{fig:deltau_r_theta}
corresponding to $R=3.5\sigma$ and $R=4.5\sigma$ ). These results
show that in a range of salt concentrations, if we consider two
rod-like polyelectrolytes in the solution at a separation where
their effective interaction is appreciable, the interaction
potential tends to reorient them to a perpendicular configuration.
The preferred orientation is parallel only at very close
separations. This could be the origin of the formation of
scaffold-like structures in the aggregation of stiff
polyelectrolytes. After the formation of such structures, the
polyelectrolytes are much closer to each other, and will have a
chance to discover that the parallel configuration has a lower free
energy, so that perpendicular rods eventually reorient themselves
and form bundles.

To elaborate further on this issue, we calculate the
distance-dependent potential of mean force between two perpendicular
rods, which shows that in the presence of 3-valent salt there can be
attractive interaction between them at close separations (see Fig.
\ref{fig:deltau_r_perpendicular}). Moreover, Fig.
\ref{fig:deltau_r_perpendicular} shows that a barrier exists in the
potential of mean force, which decreases with increasing salt
concentration. In Ref. \cite{Fazli}, it has been shown that the
perpendicular configuration of two rods when their ends are close to
each other has a lower free energy than the configuration shown in
Fig. \ref{fig:deltau_r_perpendicular}. In fact, within the dominant
kinetic mode of aggregation, when two polyelectrolytes are going to
join each other with their ends and at right angle, they experience
the lowest possible energy barrier. Using the simulation results
presented in Figs. \ref{fig:deltau_r_theta},
\ref{fig:deltau_r_parallel}, and \ref{fig:deltau_r_perpendicular},
and similar studies at other salt concentrations (not presented
here), we have calculated the preferred crossing angle, $\theta_p$,
and the interaction type between two rods at different salt
concentrations and center-to-center separations as shown in Fig.
\ref{fig:c31_r}. We find that with increasing salt concentration,
the $\theta=\frac{\pi}{2}$ region of the diagram becomes smaller,
until it finally vanishes at $c_{3:1}\simeq1.0$. Moreover, there is
a region in the $c_{3:1}-R$ plane for which the preferred angle
between the rods is $90^\circ$ and the interaction is attractive.
The existence of such a regime could explain the formation of the
scaffold-like structures during the aggregation of polyelectrolytes.
We note that the information extracted from the potential of mean
force between a pair of rods on the preferred angle between the rods
and whether the interaction is attractive or repulsive does not
provide a complete description of the behavior of the system, as the
true phase behavior needs to be studied from the global stability
(convexity) of the free energy of the system (see e.g. Ref.
\cite{harreis} for such a study in the context of columnar
DNA-aggregates). Such a study, however, will involve making
assumptions on the structure of the aggregate as studying the full
infinite dimensional configuration space of a system of $N_p$ rods
is not computationally feasible. Instead, we prefer to use the
calculation of the potential of the mean force (which is more
tractable) to supplement the results obtained from MD simulations
that probe the kinetics of the system while relaxing to equilibrium.

We also calculate the ensemble average of interaction energy
(Lennard-Jones + Coulomb) and the potential of mean force (by
integration of the torque with respect to the angle) as functions of
the angle between rods with monovalent counterions and no added
salt. Potential of mean force $\Delta U$ is a partial free energy
(subject to the constraint that the rods are kept fixed) and can be
decomposed into an energetic and an entropic part, namely $\Delta
U=\Delta E-T \Delta S$, where the energetic contribution is the
average interaction energy $\Delta E$. The potential of mean force
$\Delta U$ and the average of interaction energy $\Delta E$ are
shown in Fig. \ref{fig:deltae_deltau_theta}. The comparison shows
that although the right angle configuration is the preferred form
judging from the potential of mean force, the parallel configuration
has a lower interaction energy. This could be understood by noting
that the counterions in the parallel configuration are effectively
confined in a smaller volume around the rods than in the
perpendicular configuration, which causes an entropic penalty. This
suggests that the entropy of the counterions has a crucial role in
determining the equilibrium configuration of the system.

\section{Stability of the raft-like structures}\label{sec:Stb}

The formation of scaffold-like meta-stable structures during the
aggregation of rod-like polyelectrolytes has similarities to the
raft-like structures observed experimentally in solutions of F-actin
rods with multivalent salt \cite{G.Wong}. Since we observe that in
an intermediate range of salt concentrations the system is likely to
be trapped in meta-stable states on its way to equilibrium, the
question naturally arises as to how this finding might have
relevance to experiments on highly charged rod-like polyelectrolytes
such as F-actin in the presence of multivalent salt. The formation
of aggregates from single filaments distributed randomly in the
solution as the initial configuration is controlled by a time scale
that has two components, namely the time that takes the filaments to
find each other through diffusion and the time that takes to
overcome energetic barriers \cite{Fazli}. Since the diffusion
component of the characteristic relaxation time could be substantial
for dilute solutions, having long relaxation times does not
necessarily mean that large energetic barriers impede the formation
of equilibrium structures \cite{Fazli,gerard-ramin}. To eliminate
the diffusion component of the aggregation time and focus
specifically on the energetic barriers in the dynamics of the
system, we can study the evolution of already made complexes towards
equilibrium. While strictly speaking this corresponds to a different
kinetic pathway, one can imagine that a similar process of
disentanglement of the interacting filaments will be followed in the
aggregation kinetics, on the way to equilibrium. Therefore,
measuring the relaxations times of already formed structures could
provide a direct quantitative measure of the kinetics of the
meta-stable state of the system. To this end, we study
systematically the stability and time evolution of raft-like
aggregates under various conditions.

\begin{figure}[ht]
\begin{center}

\resizebox{1.99\columnwidth}{!}{
\includegraphics{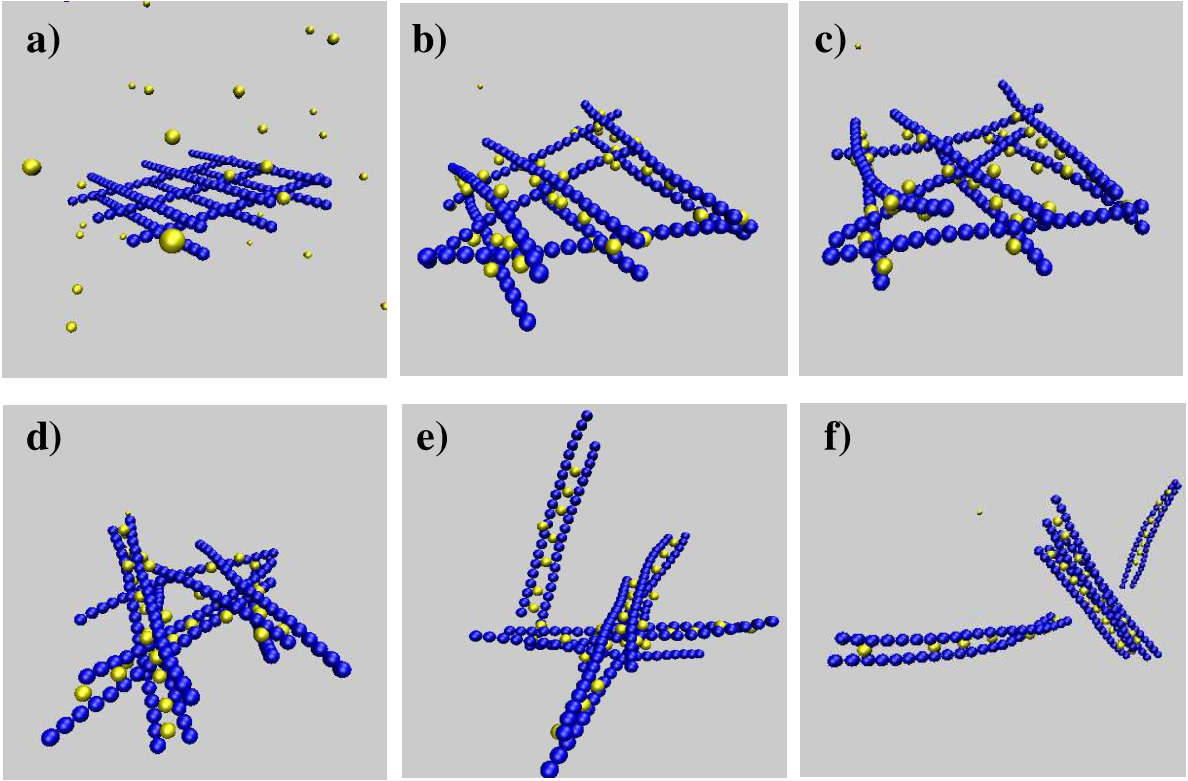}}
\caption{Snapshots from evolution of a raft-like structure formed by
$N_p=9$ polyelectrolytes in the presence of 3-valent salt of
concentration $c_{3:1}=0.5$. The positive 3-valent salt ions are
shown by golden (light) spheres and the negative salt ions and
counterions are not shown for clarity. Corresponding times of
snapshots in units of $\tau_0$: (a) initial configuration before
equilibration of the ions, (b) 50, (c) 250, (d) 400, (e) 550, (f)
700.} \label{fig:snapshot2}
\end{center}
\end{figure}

We construct raft-like structures of stiff polyelectrolytes at
different salt concentrations, and fix them at the beginning of the
simulation (see Fig. \ref{fig:snapshot2} a) while the counterions
and the salt ions are allowed to fluctuate. After the equilibration
of the free ions, we release the constraints on the polyelectrolyte
rods and follow the evolution of the system. We find that in the low
salt concentration regime thermal fluctuations destroy these
structures in a short time. In fact, in this regime, there is not
enough salt in the system for like-charge attraction between
polyelectrolytes to appear and the electrostatic repulsion between
the rods make such structures unstable. In the high salt
concentration regime also the raft-like structures have a fast
evolution and considerably short life time. In this regime mutually
perpendicular rods slide against each other in a short time and the
structure changes to bundles of parallel rods.

In the regime with intermediate salt concentration, however, the
system has a different evolution. In this regime we find that the
raft-like structures are considerably long-lived and it takes a long
time for the fluctuations to destroy them. The system finally finds
the lower free energy state in which bundles of parallel
polyelectrolytes are formed. In this regime, the stability of
raft-like structures increases with increasing salt concentration. A
time lapse view of the system with $N_p=9$ rods is shown in Fig.
\ref{fig:snapshot2}.

To help quantify the stability of the raft-like structures, we study
the dependence of the evolution time of the structures towards
bundles of parallel polyelectrolytes, as a function of the size of
these structures while keeping the volume density of the filaments
constant (by increasing the size of the simulation box with
increasing number of filaments). The evolution time $t_e$ is defined
as the time it takes for a raft-like structure containing $N_p$
polyelectrolytes to change to bundles of polyelectrolytes; for
example the time interval between snapshots a and f of Fig.
\ref{fig:snapshot2}. The evolution time $t_e$ as a function of the
number of polyelectrolytes in the raft-like structure $N_p$ is shown
in Fig. \ref{fig:te_np}a. The data points in this figure are
obtained by averaging over five different runs of the system. In
this figure, $N_p=9$ corresponds to three mutually perpendicular
layers and each layer contains three parallel rods (see Fig.
\ref{fig:snapshot2}). Similarly, $N_p=15,20,25,30,40$ respectively
correspond to $3,4,5,6$ and $8$ layers and each layer contains five
parallel rods.

The characteristic evolution time $t_e$ can be thought of as the
time it takes for the raft-like structure to disentangle the
filaments that strongly interact with each other via frustrated
electrostatic interactions, and as can be seen in Fig.
\ref{fig:te_np}a it increases with increasing the size of the
raft-like structure $N_p$. The increase is faster than linear, and
can be approximated as a power law with an exponent of about $2.7$
(see the inset of Fig. \ref{fig:te_np}a for a log-log plot of $t_e$
versus $N_p$), although the numerical data are not over a
sufficiently large domain so that the power law nature can be
verified. The characteristic relaxation time of the system as a
function of the number of filaments (for a fixed volume density of
single filaments) can be used to estimate the $N_p$ dependence of
the energy barrier (or activation energy) $E_a$ for the transition
between the raft-like structure and the equilibrium structure.
Assuming a form of $t_e(N_p)=t_0 e^{E_a(N_p)/k_{\rm B} T}$, where
$t_0$ is a microscopic time scale that does not depend on the number
of filaments, we can calculate relative changes in the activation
energy as shown in Fig. \ref{fig:te_np}b. While the plot in Fig.
\ref{fig:te_np}b clearly shows that the energy barrier increases
with the number of filaments, more studies are necessary to
determine whether this trend will continue or ultimately level off
for systems with considerably larger number of filaments.

Using the estimate for the microscopic time scale $\tau_0=1.3$ ps,
we find that for 9 rods the relaxation time is $\sim 1$ ns and for
40 rods it is $\sim 0.1$ $\mu$s. To make a crude estimate for the
typical relaxation times corresponding to realistic experimental
cases \cite{G.Wong}, we can extrapolate the empirical scaling law to
$N_p \sim 10^7$, which yields $t_e\sim 10^7$ s. This estimate
suggests that the time the system could be trapped in a meta-stable
state could be considerably long compared to typical observations
times, for a system like F-actin. Further studies are necessary to
determine whether the super-linear increase of the relaxation time
as a function of the number of filaments persists for such large
numbers present in the experiments on F-actin.

\section{Concluding Remarks}\label{sec:Con}

We have studied the aggregation of stiff polyelectrolytes in bulk
solution in the presence of multivalent salt, and have observed
spontaneous formation of orientationally ordered aggregates. Our
results show that in a range of salt concentrations, when two
polyelectrolyte rods are at separations where their effective
interaction is considerable, they could prefer to be perpendicular
while their interaction is attractive. In this range of salt
concentrations, we have observed the formation of relatively
long-lived scaffold-like structures of the polyelectrolytes in the
bulk solution, for which the mode of evolution through to the
bundles of parallel polyelectrolytes is considerably slower than the
typical kinetic modes of the system. We have also observed that in
the above-mentioned range of the salt concentrations already formed
raft-like structures similar to the ones observed in Ref.
\cite{G.Wong} have longer life time than those in the high and low
salt concentration regimes.

Our findings suggest that when multivalent salt concentration is
enough for like-charge attraction between polyelectrolytes but is
not so high that they join up in parallel configuration directly,
formation of raft-like structures could be possible as a meta-stable
state. In our model system, which contains a small number of
polyelectrolytes, the transition from the meta-stable state to the
lower free energy state of bundles of parallel polyelectrolytes
occur over a finite observable time. This evolution time, however,
is observed to increase rapidly with the system size. This means
that in a real experimental system with a large number
polyelectrolytes involved in such a raft-like structure, it will be
extremely difficult for the system to find the state with lower free
energy in which the polyelectrolytes form dense bundles of parallel
polyelectrolytes. In other words, our studies suggest that the
experimentally observed raft-like structures might be meta-stable
states with life times that are far larger than the experimentally
available observation time. Further studies on larger systems and
over longer simulation times are needed to fully verify this
proposition, and characterize the nature of the intermediate
scaffold-like regime.

%\acknowledgements

It is a great pleasure to acknowledge stimulating discussions with
Gerard Wong.

%The references should start on their own page.
%\vspace{1.5cm}
%\noindent \textbf{Author Name 1,$^a$ Author 2 Name$^b$ and Author 3 Name$^a$$^b$}\\
%\noindent $^a$ \textsl{Address, Address, Town, Country. E-mail: xxxx@aaa.bbb.ccc\\
%\noindent $^b$ Address, Address, Town, Country. E-mail:
%xxxx@aaa.bbb.ccc}

\begin{figure}[ht]
\begin{center}
\resizebox{1.99\columnwidth}{!}{
\includegraphics{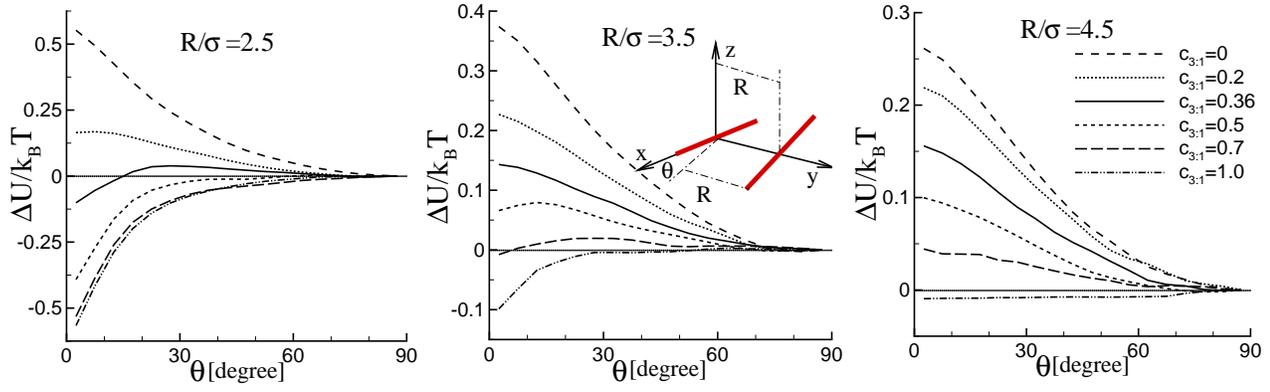}}
\caption{Potential of mean force between two rods as a function of
$\theta$ at different values of $R$ and salt concentration $c_{3:1}$
(see the schematic configuration of the rods in the inset of the
middle plot). The legends of the other two plots are the same as
that of the rightmost plot.} \label{fig:deltau_r_theta}
\end{center}
\end{figure}

\begin{figure}[ht]
\begin{center}
\includegraphics{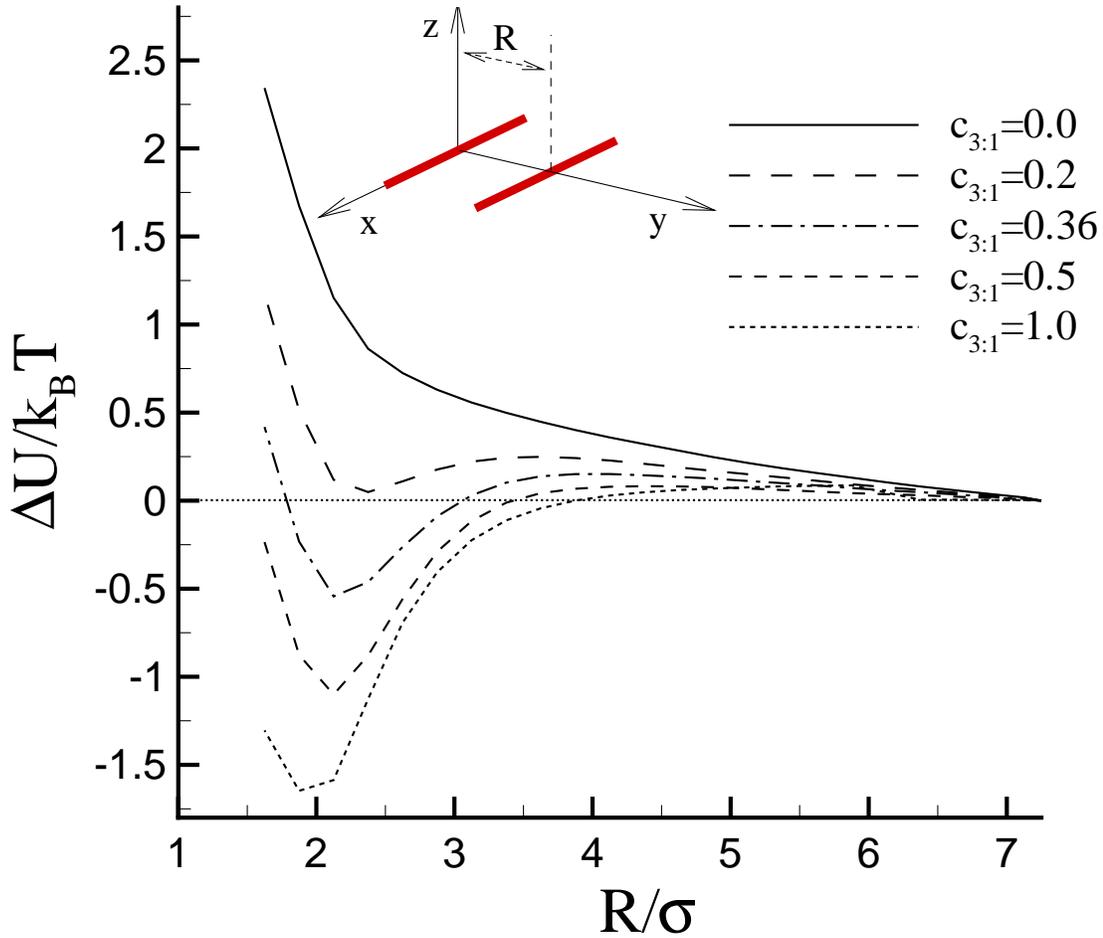}
\caption{Potential of mean force between two parallel rods (see the
schematic configuration of the rods in the inset) as a function of
center-to-center separation $R$ for different concentrations of
3-valent salt. It can be seen that at salt concentrations
$c_{3:1}\geq 0.2$ there is attraction between two parallel
polyelectrolytes.} \label{fig:deltau_r_parallel}
\end{center}
\end{figure}

\begin{figure}[ht]
\begin{center}
\includegraphics{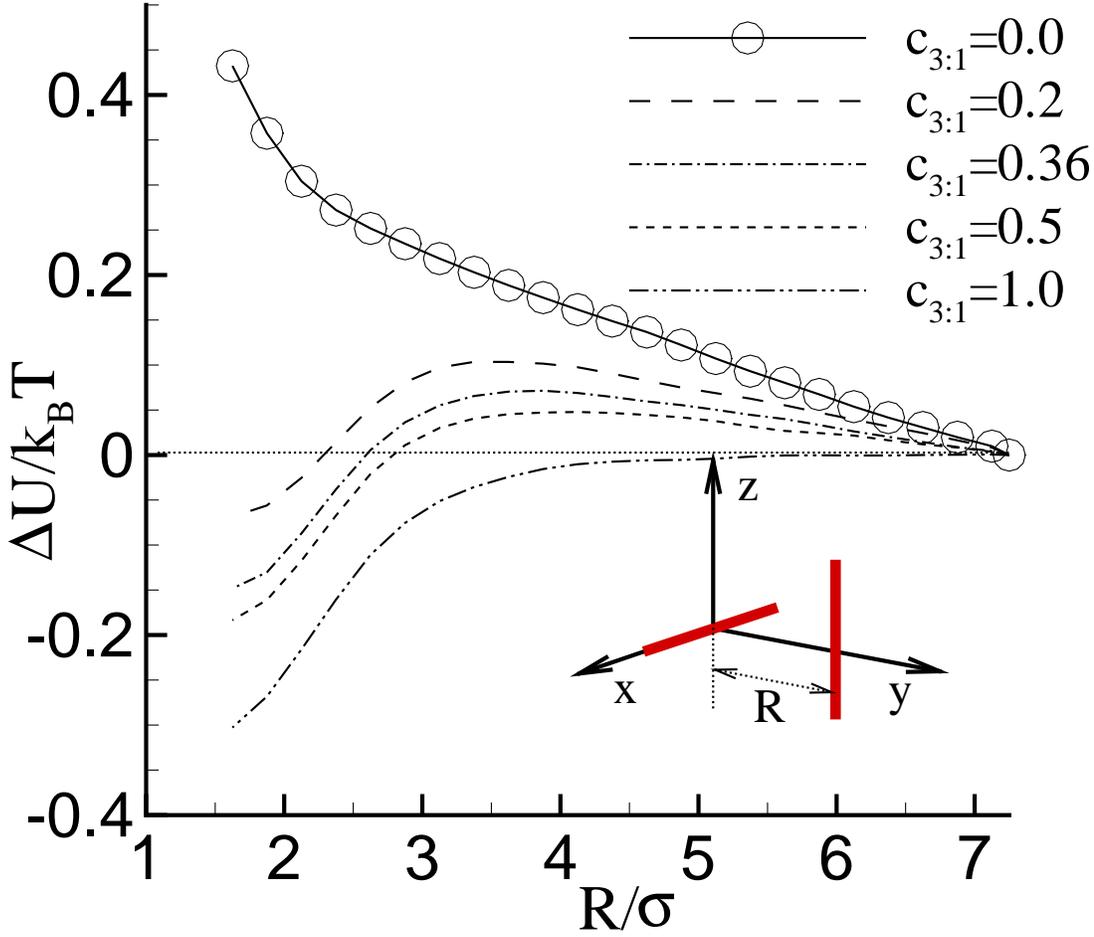}
\caption{Potential of mean force between two perpendicular rods
(schematic configuration of the rods is shown in the inset) as a
function of $R$ at different concentrations of 3-valent salt. The
size of the circles on the $c_{3:1}=0$ curve shows the size of error
bars for all our data points. Note that the potential is not shown
for $R < 1.6 \sigma$ where the excluded volume interaction takes
over and generates a strong repulsion.}
\label{fig:deltau_r_perpendicular}
\end{center}
\end{figure}

\begin{figure}[ht]
\begin{center}
\includegraphics{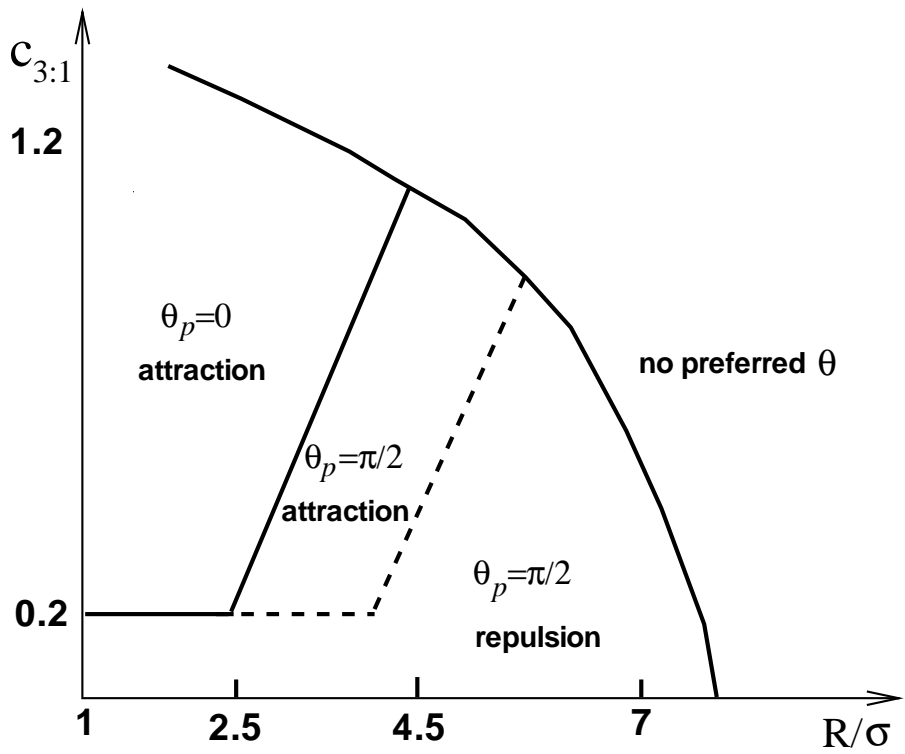}
\caption{Preferred angle $\theta_p$ and interaction type between
two charged rigid rods which are fixed on two parallel plates of
separation $R$, in different regions of $c_{3:1}-R$ plane (see the
schematic configuration of the rods in the middle plot of Fig.
\ref{fig:deltau_r_theta}).}
\label{fig:c31_r}
\end{center}
\end{figure}

\begin{figure}[ht]
\begin{center}
\includegraphics{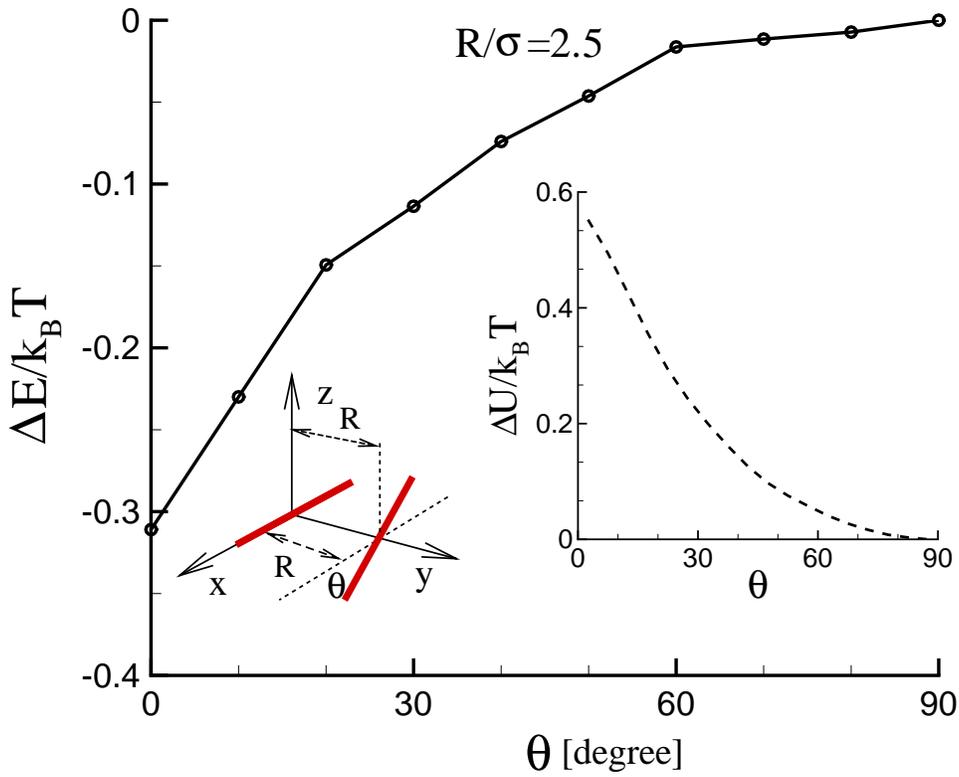}
\caption{Ensemble average of the interaction energy (Lennard-Jones +
Coulomb) per particle between two rods with their separation fixed
at $R=2.5\sigma$ as a function of $\theta$ with no added salt.
Inset: potential of mean force between the same two rods. The
different trends of the two curves shows that the entropy of
counterions has an important role in the orientation of the rods.}
\label{fig:deltae_deltau_theta}
\end{center}
\end{figure}

\begin{figure}[ht]
\begin{center}
\resizebox{1.99\columnwidth}{!}{
\includegraphics{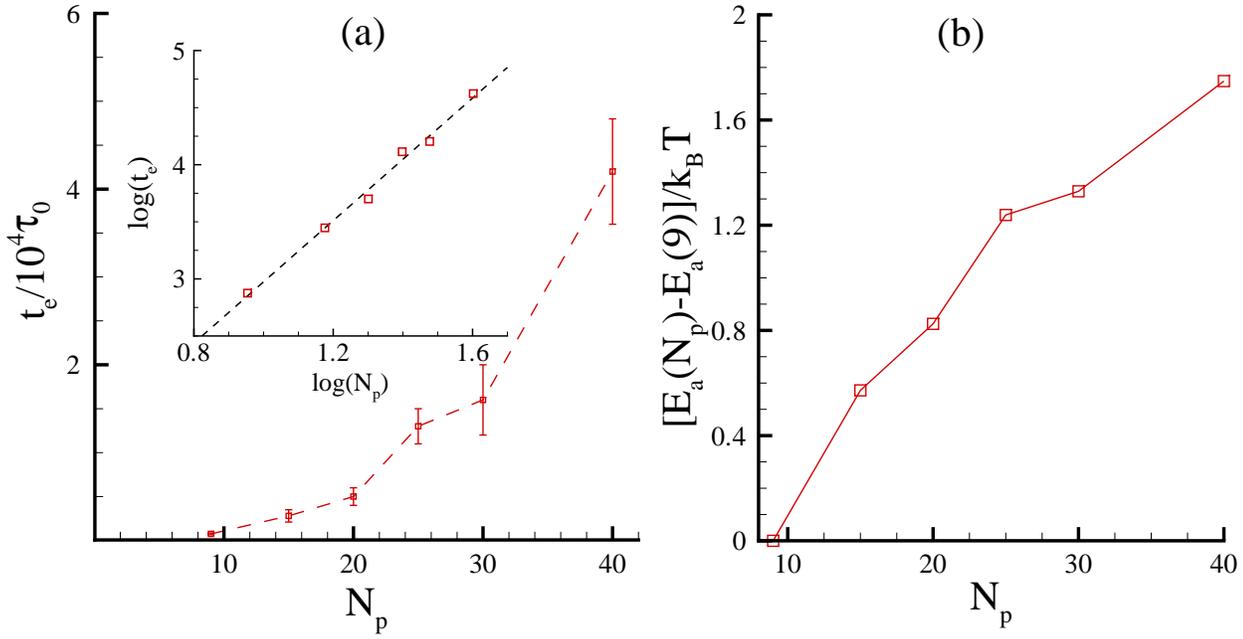}}
\caption{(a) Evolution time of raft-like structures to bundles of
parallel polyelectrolytes as a function of $N_p$, number of
polyelectrolytes forming these structures. Error bars are obtained
from averaging over five runs of the simulation with different
random number sequences. Inset: Log-log plot of $t_e$ versus $N_p$
and a line with the slope of $2.7$ (dashed line). (b) Activation
energy extracted from (a) as a function of $N_p$.} \label{fig:te_np}
\end{center}
\end{figure}

%Please compile a list of all figure captions on a separate page:
\clearpage
\begin{list}{}{\leftmargin 2cm \labelwidth 1.5cm \labelsep 0.5cm}

\item[\bf Fig. 1] Snapshots of the system containing $N_p=9$ polyelectrolytes
in solution with 3-valent salt concentration $c_{3:1}=0.5$ as a
function of time. +3 salt ions are shown by golden (light) spheres,
+1 counter-ions by red (dark), and -1 salt ions by gray spheres.
Corresponding times of snapshots in units of $\tau_0$: (a) 10, (b)
25, (c) 40, (d) 55, (e) 70, (f) 85. Aggregates of polyelectrolytes
with scaffold-like structures can be seen in the snapshots e and f.
Finally, all of the polyelectrolytes form a single bundle (last
configuration is not shown here).

\item[\bf Fig. 2] Snapshots from evolution of a raft-like
structure formed by $N_p=9$ polyelectrolytes in the presence of
3-valent salt of concentration $c_{3:1}=0.5$. The positive 3-valent
salt ions are shown by golden (light) spheres and the negative salt
ions and counterions are not shown for clarity. Corresponding times
of snapshots in units of $\tau_0$: (a) initial configuration before
equilibration of the ions, (b) 50, (c) 250, (d) 400, (e) 550, (f)
700.

\item[\bf Fig. 3] Potential of mean force between two
rods as a function of $\theta$ at different values of $R$ and salt
concentration $c_{3:1}$ (see the schematic configuration of the rods
in the inset of the middle plot). The legends of the other two plots
are the same as that of the rightmost plot.

\item[\bf Fig. 4] Potential of mean force between two
parallel rods (see the schematic configuration of the rods in the
inset) as a function of center-to-center separation $R$ for
different concentrations of 3-valent salt. It can be seen that at
salt concentrations $c_{3:1}\geq 0.2$ there is attraction between
two parallel polyelectrolytes.

\item[\bf Fig. 5] Potential of mean force between two
perpendicular rods (schematic configuration of the rods is shown in
the inset) as a function of $R$ at different concentrations of
3-valent salt. The size of the circles on the $c_{3:1}=0$ curve
shows the size of error bars for all our data points. Note that the
potential is not shown for $R < 1.6 \sigma$ where the excluded
volume interaction takes over and generates a strong repulsion.

\item[\bf Fig. 6] Preferred angle $\theta_p$ and interaction type between
two charged rigid rods which are fixed on two parallel plates of
separation $R$, in different regions of $c_{3:1}-R$ plane (see the
schematic configuration of the rods in the middle plot of Fig.
\ref{fig:deltau_r_theta}).

\item[\bf Fig. 7] Ensemble average of the interaction energy
(Lennard-Jones + Coulomb) per particle between two rods with their
separation fixed at $R=2.5\sigma$ as a function of $\theta$ with no
added salt. Inset: potential of mean force between the same two
rods. The different trends of the two curves shows that the entropy
of counterions has an important role in the orientation of the rods.

\item[\bf Fig. 8] (a) Evolution time of raft-like structures to
bundles of parallel polyelectrolytes as a function of $N_p$, number
of polyelectrolytes forming these structures. Error bars are
obtained from averaging over five runs of the simulation with
different random number sequences. Inset: Log-log plot of $t_e$
versus $N_p$ and a line with the slope of $2.7$ (dashed line). (b)
Activation energy extracted from (a) as a function of $N_p$.

\end{list}

\clearpage

\clearpage

\clearpage

\clearpage

\clearpage

\clearpage

\clearpage

\clearpage

\end{document}